\documentclass[fleqn,twoside]{article}
\usepackage{espcrc2}
\usepackage{epsfig}
\usepackage[figuresright]{rotating}

\newcommand{\AmS}{{\protect\the\textfont2
  A\kern-.1667em\lower.5ex\hbox{M}\kern-.125emS}}
\title{CNGS neutrino beam systematics for $\theta_{13}$}

\author{A. Ferrari	
    \address{CERN, 1211 Geneve 23, Switzerland.}\thanks{On leave of absence from INFN Sez. di 
Milano.}
    A. Guglielmi
    \address{Istituto Nazionale di Fisica Nucleare and Dept. of Physics, via Marzolo 8, 35131 
            Padova, Italy.}
	P. R. Sala
       \address{Istituto Nazionale di Fisica Nucleare, via Celoria 16, 20133 Milano, Italy.}}
        
\begin{document}

\begin{abstract}
Energy spectra, intensity and composition of the CERN to Gran Sasso CNGS neutrino beam for 
$\nu_\mu \rightarrow \nu_\tau$ and $\nu_\mu \rightarrow \nu_e$ oscillation searches are presented.
The associated beam systematics, which is  the major ingredient for the  $\nu_\mu \rightarrow 
\nu_e$ search sensitivity,  are obtained from the study of the previous CERN WANF.
\vspace{1pc}
\end{abstract}

\maketitle

\section{Introduction}

 Over the next five years the present generation of oscillation experiments at
 accelerators with long-baseline $\nu_{\mu}$ beams,
 K2K at KEK \cite{K2K}, MINOS  at NUMI beam of FNAL \cite{NUMI} and ICARUS and 
 OPERA \cite{ICARUS-OPERA} at CNGS CERN to Gran Sasso neutrino beams \cite{CNGS} is expected
 to confirm the $\nu_{\mu} \rightarrow \nu_{\tau}$ transitions observed 
 in the atmospheric $\nu$  and measure $\sin^2 2 \theta_{23}$ and
 $|\Delta m^2_{23}|$ within $10 \; \%$  of accuracy if
 $|\Delta m^2_{23}| >  10^{-3}$ eV$^2$.
 K2K and MINOS are looking for neutrino disappearance, by measuring
 the $\nu_{\mu}$ survival probability as a function of neutrino energy while
 ICARUS and OPERA will search for evidence
 of $\nu_{\tau}$ interactions in   conventional $\nu_{\mu}$ beams.

 Even not explicitly optimized for $\nu_{\mu} \rightarrow \nu_e$ searches
 these experiments  can explore  $\sin^2 2 \theta_{13}$ 
 oscillation parameter beyond the CHOOZ limit  \cite{CHOOZ}. 
 In particular ICARUS and OPERA \cite{ICARUS-OPERA} 
 can reach a combined sensitivity on $\sin^2 2 \theta_{13}$ 
 (CP violation and matter effects not accounted for) a factor $\sim 5$ better than CHOOZ
 in the allowed parameter region of atmospheric neutrino oscillations for five years 
 exposure to the CNGS beam at nominal intensity  \cite{migliozzi1}.
 The $\nu_{\mu} \rightarrow \nu_e$  oscillations will be searched for as $\nu_e$ charge current events
 excess over the $\nu_e$ contamination of the beam: the key issue will be the knowledge of the neutrino beam
 composition and spectrum. A strong impact on $\sin^2 2 \theta_{13}$ sensitivity is expected
 from the systematics of the $\nu_e/\nu_\mu$ ratio, and in particular from its normalization error.

  The CNGS beam-line design  was accomplished 
  on the basis of the previous WANF $\nu_\mu$ beam experience at CERN SPS  \cite{wanf1} for 
  CHORUS and NOMAD experiments \cite{CHORUS-NOMAD} which allowed for a powerful study 
  of the neutrino beam  providing  a strong benchmark for conventional neutrino beams and 
  in particular for the CNGS. In this sense the CNGS beam systematics on $\nu_e/\nu_\mu$ ratio
  can be derived and predicted from the previous measurements and studies performed with the WANF.

 \section{The WANF: a case of study}

 In the WANF $\nu_\mu$ facility 450 GeV/c protons were extracted by the SPS and sent into 
 a segmented target composed by 11 Be rods, 10 cm of length and 3 mm diameter each. 
  The produced positive (negative) mesons, essentially 
 $\pi$ and $K$, were focused (defocused) by  two pulsed magnetic lenses (horn and reflector) 
 into a 290 m vacuum decay tunnel were neutrinos  were produced. The resulting $\nu_\mu$  beam
 in NOMAD at 840 m from the Be target, was characterized by an energy $\sim 24.3$ GeV and
 a contamination of $\sim 6.8 \%$, $\sim 1 \%$ and $0.3 \%$ of $\overline{\nu}_\mu$, $\nu_e$ and 
 $\overline{\nu}_e$ respectively.
  Four main processes as sources of neutrinos were recognized:
 
 \begin{itemize}
   \item 450 GeV/c proton interactions in the Be target which produce $\pi^{\pm}$,
         $K^{\pm}$, $K^0$,... originating $96 \%$ of the $\nu_\mu$ flux in NOMAD;
	 
   \item primary protons non interacting or missing the target which produce
         in the horn and reflector walls, 
         windows and beam-dump, mesons only weakly or not focused by the
         beam-optics ($\sim 15 \%$ of $\overline{\nu}_\nu + \overline{\nu}_e$);
	 
   \item ``prompt neutrinos'' generated in the decay of charmed particles and kaons
         in the target and dump yielding $\sim 6 \%$ of $\overline{\nu}_e$;
	 
   \item particles reinteractions along the beam-line affecting $\sim 10 \%$ of 
         $\nu_\mu$ and $\nu_e$ but $\sim 40 \%$ of defocused component 
	 $\overline{\nu}_\mu + \overline{\nu}_e$.
                                                                                                          
 \end{itemize}

\begin{figure}[!htp]
  \begin{center}
  \vskip -1.5cm
    \mbox{\hspace*{-1.0cm}\epsfig{file=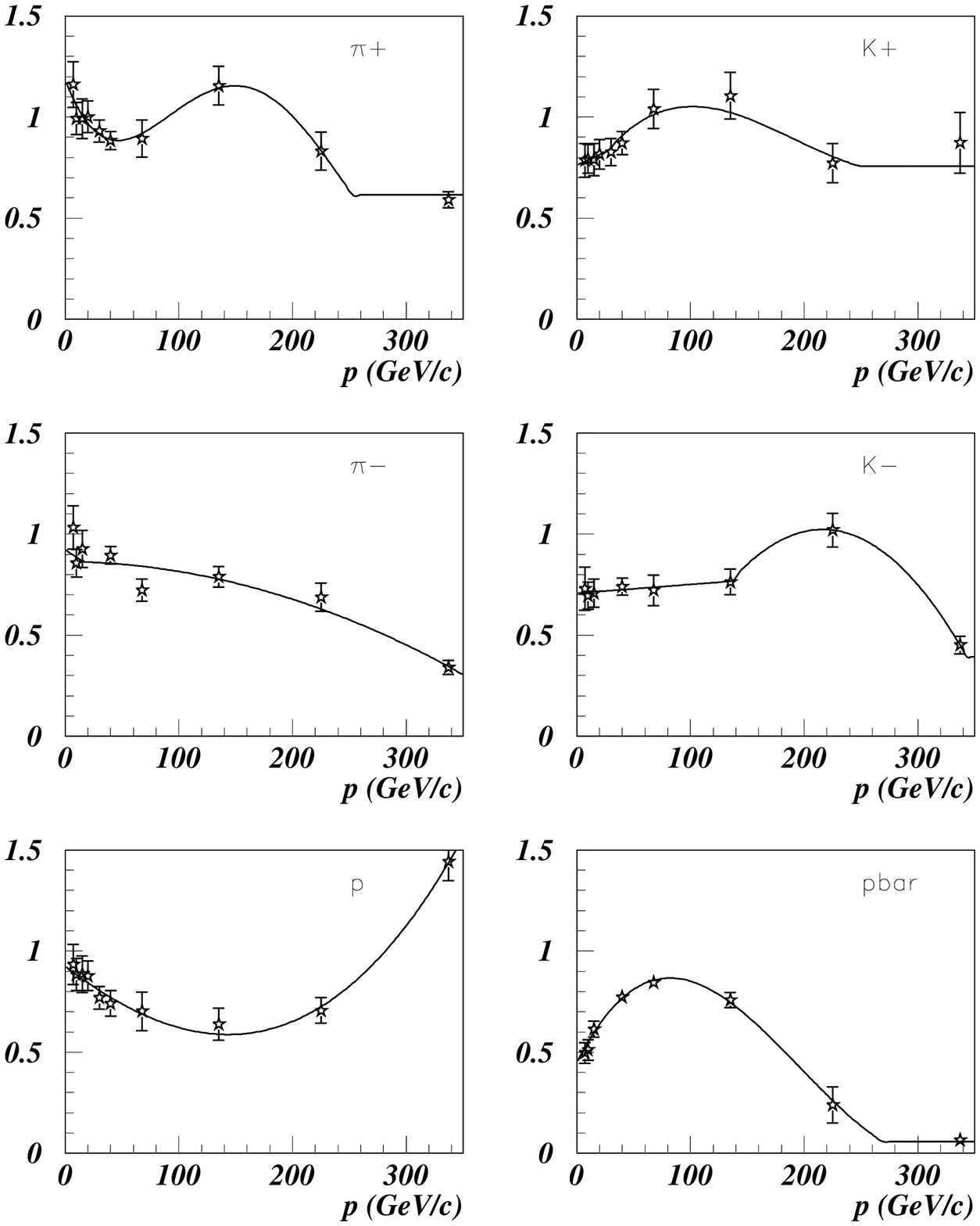,width=9.5cm,height=7.3cm}}
   \vskip -0.5cm
  \mbox{\hspace*{-1.0cm}\epsfig{file=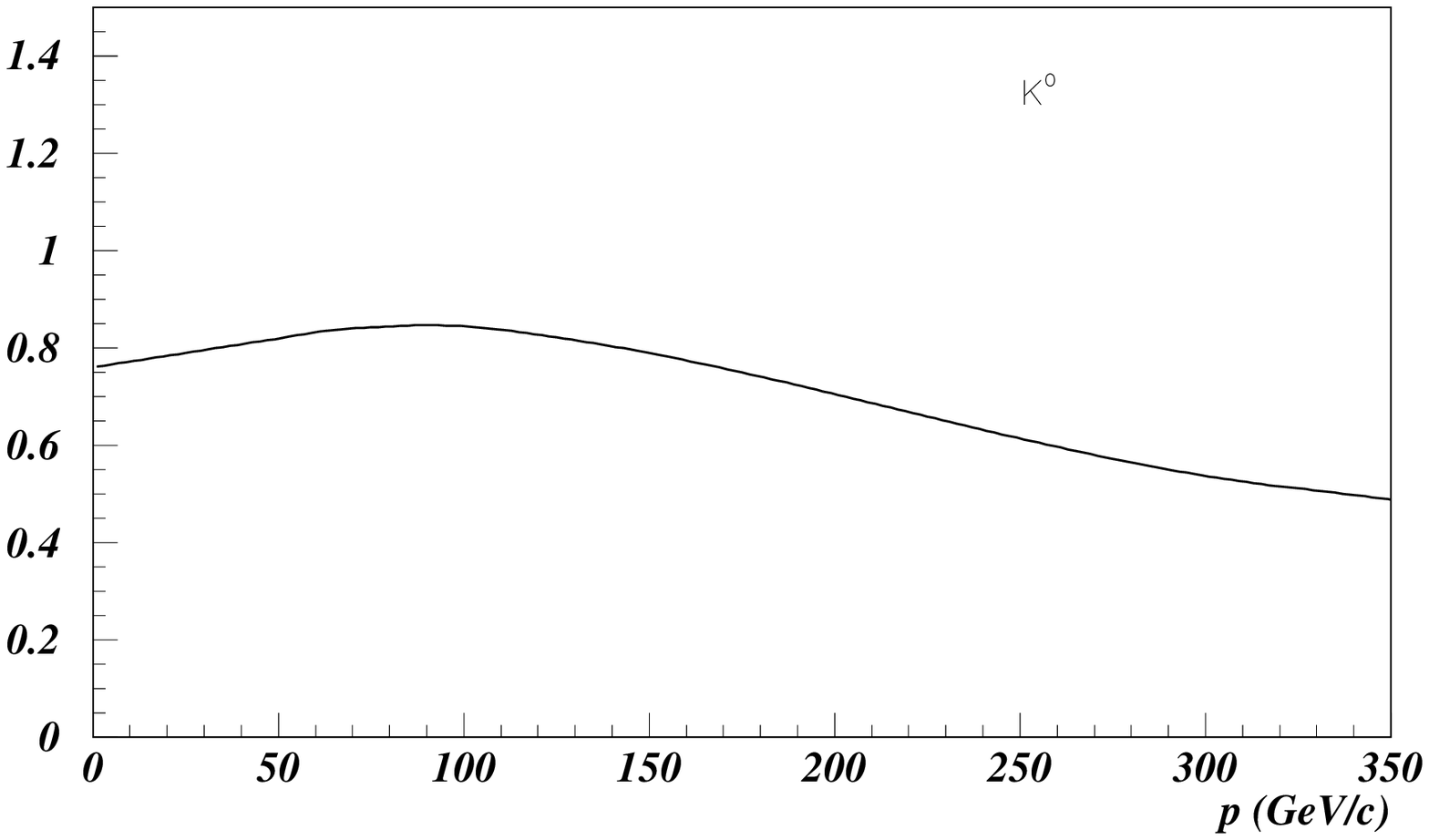,width=9.5cm,height=7.3cm}}
  \end{center}
  \vskip -5.0cm
  \caption{FLUKA to SPY, NA20 meson production corrections as a function of the momentum $p$.
           The curves are the results of fitting the weights at fixed $p$ (showed with their 
           uncertainty bars) with combination of polynomial functions \cite{wanf2}.}
  \label{corr}
 \end{figure}

 \begin{figure}[!hbt]
  \begin{center}
  \vspace*{-0.8cm}
  \mbox{\epsfig{file=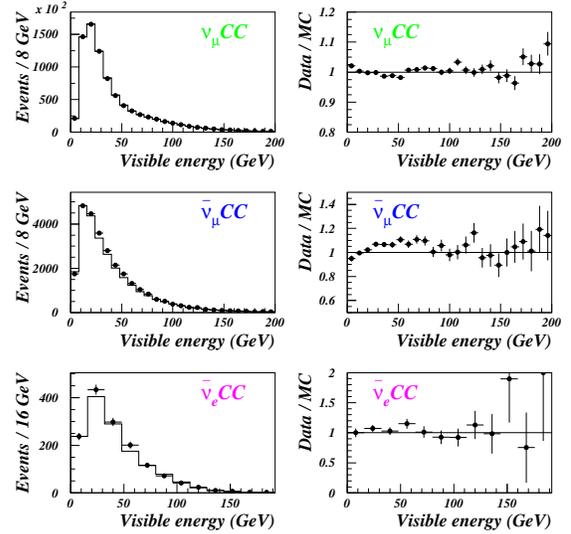, width=8cm, height=8.0cm}}
  \end{center}
  \vspace*{-1.5cm}
 \caption{Neutrino energy spectra (left) for the data (points
                with statistical error bars) and the Monte Carlo (histogram) for
                $\nu_{\mu}$ CC, $\overline{\nu}_{\mu}$ CC and
                $\overline{\nu}_e$ CC interactions 
                and their corresponding ratios (right) in
                NOMAD \cite{wanf2}.}
  \label{nu-CC}
 \end{figure}

\begin{figure}[!hb]
  \begin{center}
  \vspace*{-1.0cm}
  \mbox{\epsfig{file=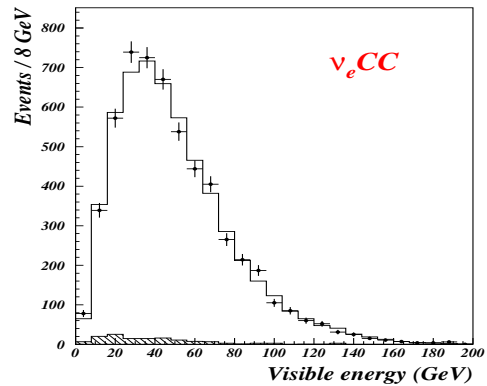, width=7cm, height=5.6cm}}
  \end{center}
  \vskip -1.3cm
  \caption{Energy spectrum  for the measured  $\nu_e$ CC  interactions
                (points with statistical error bars) and the Monte Carlo (histogram) in
		NOMAD \cite{wanf2}.}
  \label{nue-CC}
 \end{figure}

\begin{figure*}[!htp]
  \begin{center}
  \mbox{\epsfig{file=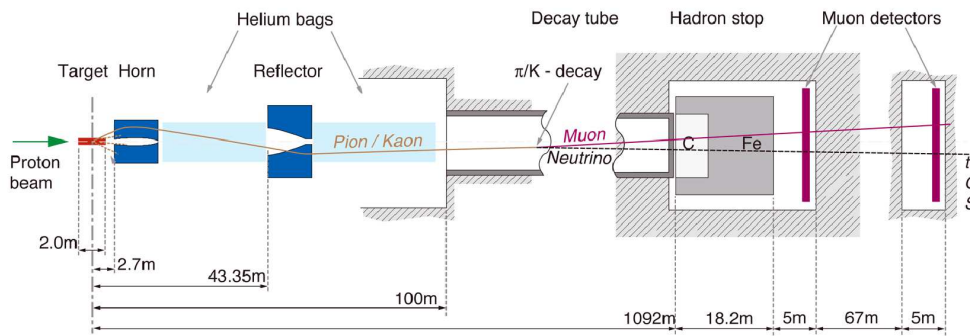,angle=-90,width=13.0cm}}
  \end{center}
  \vskip -0.5cm
  \caption{Schematic layout of the future CNGS neutrino beam line.}
  \label{cngslayout}
 \end{figure*}

\vskip -0.7cm
 In order to predict energy spectra, intensity and composition of the $\nu$ beam a
 sound knowledge  of meson production in the target is required. 
 Indeed the accurate description of the $K^+$ and $K^0$
 relative  to the $\pi^+$ is essential to calculate the $\nu_e$ 
 initial content in the beam.

 \noindent Direct measurements of $\pi$ and $K$ production in Be 
 by 400 GeV/c protons were performed by NA20 Collaboration \cite{na20} and
 by the SPY Collaboration \cite{spy} at 450 GeV/c.
  Accuracy of Monte Carlo generators of hadronic interactions for
 the meson production could limit the sensitivity to neutrino 
 oscillation searches.
 The best agreement between predictions and data was found with FLUKA
 standalone code \cite{FLUKA} which reproduces the measured $\pi$ and $K$ 
 yields at the level of $\sim 20 \%$ in the momentum region $ 30 \div 100$ GeV/c 
 which is expected to contribute most to  $\nu$ flux  \cite{generators}. 
 Furthermore an accurate description of the primary proton beam spot,
 focusing system as well as of the materials inserted in the beam line 
 from the target to the dump (hadronic reinteraction processes) 
 is also mandatory especially for the neutrino beam minor components.

  A complete analysis of this $\nu$ beam  was performed by a 
 beam line simulation from the Be target up to the NOMAD 
 detector including the FLUKA generator with further corrections to the meson
 production based on the residual differences between the 
 predicted and measured meson yield in Be.
 These reweighting functions were calculated as a function of the meson momentum by integrating 
 the particle production over 10 mrad of angular acceptance and including also the transport efficiency
 as a function of production angle along the beam-line.
 The $K_0$ reweighting was obtained from the corresponding for $K^{\pm}$  using a 
''quark-counting model''  with 15 \% of uncertainty \cite{wanf2}.
  
  \noindent The agreement with the measured $\nu$ interactions 
  in NOMAD was at the few percent level (fig. \ref{nu-CC}) with a systematic 
  normalization error of $\sim 7 \%$ on
  $\nu_{\mu}$CC/p.o.t. essentially determined by the accuracy on  the $\pi^+$ and $K^+$ production
  in the Be target ($3.4 \%$) and on the proton beam spot position on the Be target.
  The corresponding energy dependent error ranged between 2 and $5 \%$ in the $2 < E_\nu< 100$ GeV
  neutrino energy.
  Due to correlations between the sources of the $\nu_\mu$ and $\nu_e$ fluxes, the 
  normalization uncertainties on $\nu_e/\nu_{\mu}$ ratio was smaller the the uncertainties on the individual 
  $\nu_\mu$ and $\nu_e$ fluxes: $4.2 \%$ \cite{wanf2}. The corresponding energy-dependent uncertainty  
  ranged from 4 to 6 $\%$.

\begin{figure*}[!htp]
  \begin{center}
  \vskip -0.9cm
  \mbox{\epsfig{file=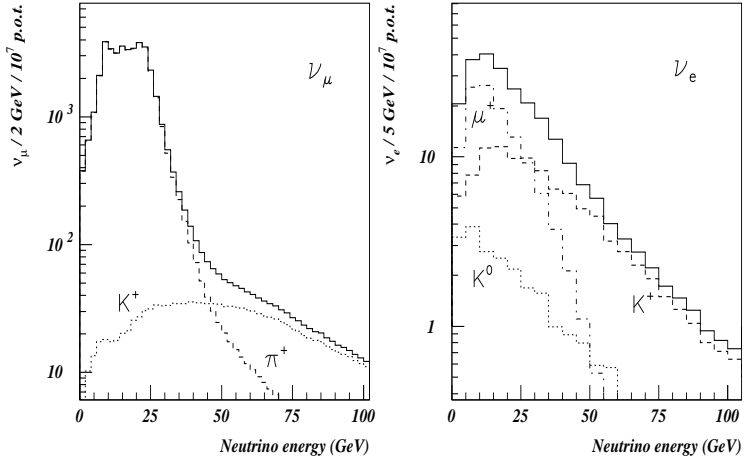,width=11.0cm,height=13.2cm}}
  \end{center}
  \vskip -7.1cm
  \caption{Muon and electron neutrino spectra with parent particles at the Gran Sasso
           site.}
  \label{cngs-beam}
 \end{figure*}

  \section{The CNGS neutrino beam}

 In the CNGS neutrino beam facility (Fig. \ref {cngslayout})
 the primary protons at 400 GeV/c of momentum with  nominal intensity of $4.5 \cdot 10^{19}$
 pot/year (proton beam shared operations),
 will be extracted from the CERN SPS and sent to a carbon target, 13 graphite rods of 10 cm 
 length, 5 and 4 mm of diameter, the first eight  spaced by 9 cm.
 Similarly to previous WANF neutrino 
 beam a magnetic horn and a reflector will allow the focusing (defocusing) of positive 
 (negative)charged secondaries
 into a $\sim$ 1 km long decay tunnel where an intense $\nu_\mu$ neutrino beam is produced.
 A large graphite and iron dump will absorb the residual hadrons at the end on the beam-line.
 A particular attention was devoted to have the full containment of the proton beam spot in the 
 target and to reduce as well as possible the quantity of material in the beam line in order
 to maximize the neutrino flux in the Gran Sasso site.

  A full detailed simulation of the CNGS beam line was performed within FLUKA framework.
  At the nominal proton beam intensity $4.5 \cdot 10^{19}$
 pot/year (proton beam shared operations) roughly 2900 $\nu_\mu$CC/kt/year are
 expected at the Gran Sasso site.  
 The increase the proton beam intensity has been studied, resulting in a
 possible upgrade bigger than $40 \%$ \cite{SPS_pot_increase}.

 The  muon neutrino flux will be characterized by  an average energy of $17.4$ GeV and 
 $\sim 0.6 \%$ $\nu_e$ to $\nu_{\mu}$ contamination
 for $E_{\nu} < 40$  GeV (Fig. \ref{cngs-beam}). The $\overline{\nu}_\mu$ and 
 $\overline{\nu}_e$ component are below $2 \%$ and $0.2 \%$ respectively.
 The $\nu_\tau$ intrinsic level in the beam will 
 be below $10^{-6}$ allowing for clean $\nu_\mu \rightarrow \nu_\tau$
 appearance experiments with both ICARUS and OPERA detectors.
 Due to the 732 Km of baseline the contribution to neutrino beam 
 from the $K^\pm$, $K^0$ is reduced by a factor $1.5\div 2$ with respect to the WANF,
 while those from high energy proton interactions downstream of the graphite target and
 from prompt charmed particles and kaons decay in the target and dump are negligible.
 The $\nu_e$ component will be mainly produced in the $\mu^+$ decay instead from $K^+$ and $K^0$ 
 as in the WANF.

 \section{CNGS beam systematics}

 The previous scheme used for the WANF neutrino
 beam calculation based on FLUKA and SPY, NA20 hadroproduction data in Be is well suitable
 for the CNGS neutrino neutrino beam.
 
 \noindent The Be-C scaling due to the different target material is expected to be
 almost independent on the secondary meson type, to depend only weakly on the transverse 
 momentum of the meson, so contributing less than $2 \%$ to the uncertainty on neutrino flux
 and less than $ 0.5 \%$ to $\nu_e/\nu_\mu$ ratio.                                                                                                        
  
 The systematics of neutrino flux at Gran Sasso can be evaluated from the WANF by properly
 rescaling the contribution of each meson to neutrino flux according to the different
 production processes.
 A total normalization error of $\sim 4 \%$  can be predicted for  both $\nu_\mu$
 and $\nu_e$ where $\sim 3.8 \%$ of uncertainty is  the secondary production
 at the target and  $\sim 1.1 \%$ error is due to particle reinteractions along
 the beam-line.
 Owing to the target configuration and the 732 km base-line, the position of proton beam on 
 the target section, the beam-line optics such as horn and reflector currents and the amount
 of material included in the simulation are not critical but contribute only $0.8 \%$ to
 normalization error of $\nu_\mu$ and $\nu_e$ flux. 
 Due to correlations between the sources of the $\nu_\mu$ and $\nu_e$ fluxes,
 where both  neutrinos are essentially produced  from $\pi^+$ of different momenta, 
 the $\nu_\mu$ by pion decay and $\nu_e$ from the subsequent muon decay,
 a large cancellation occurs in the $\nu_e/\nu_\mu$ error evaluation.
 Therefore the  normalization uncertainty on $\nu_e/\nu_\mu$ is expected to be only 
 $\sim 3.1 \%$. 
 The corresponding bin-to-bin energy dependent error is estimated to be  $2.5 \div 4 \%$ in
 the relevant part of the spectrum for $\nu_\mu$, $\nu_e$ and $\nu_e/\nu_\mu$ ratio.
  
 \begin{figure}[!htp]
  \begin{center}
  \vskip -1.0cm
  \mbox{\epsfig{file=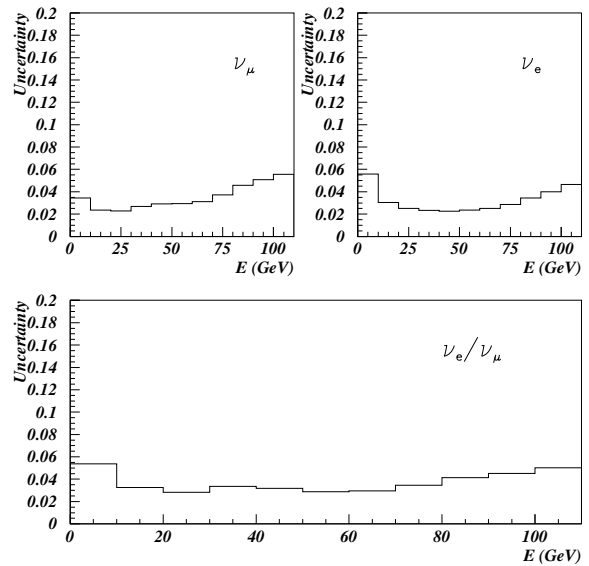,width=8.5cm,height=8.5cm}}
  \end{center}
  \vspace{-1.5cm}
  \caption{Energy dependent uncertainty for $\nu_\mu$ and $\nu_e$ neutrino
           flux and for their ratio $\nu_e/\nu_\mu$.}
  \label{numu_nue_shape}
 \end{figure}
  \vspace{-0.5cm}
 
 \section{Conclusions}
 
  The CERN to Gran Sasso CNGS neutrino beam project for $\nu_\mu \rightarrow \nu_\tau$ 
  oscillation search with ICARUS and OPERA detectors largely benefits of
  the WANF experience with CHORUS and NOMAD experiments.
  
  \noindent The neutrino flux can be predicted at Gran Sasso within a systematic uncertainty $\sim 5 \%$,
  the $\nu_e/\nu_\mu$ ratio will be known with $\sim 3.1 \%$ normalization error and $3\div 4 \%$
  energy dependent error, better than quoted by
  ICARUS and OPERA, opening the possibility to search for $\nu_\mu \rightarrow \nu_e$
  oscillation exploring $\sin^2 2 \theta_{13}$ beyond the CHOOZ limit.

   \noindent However the sensitivity on both $\nu_\mu \rightarrow \nu_\tau$ and 
   $\nu_\mu \rightarrow \nu_e$ oscillation channels are completely dominated by the statistics. 
   In 5 years of CNGS standard operations,
   $4.5 \cdot 10^{19}$ pot/year, 12 $\nu_\tau$ CC events can be recognized in ICARUS and
   similarly in OPERA if $\Delta m^2_{23} = 2.6 \cdot 10^{-3}$ eV$^2$. In the same time an 
   excess of only 21 $\nu_\mu \rightarrow \nu_e$ CC events over 120 $\nu_e$CC from beam 
   contamination and $\nu_\mu \rightarrow \nu_\tau$ is expected  in ICARUS (2.35 kton fiducial 
   mass) for $E <  20$ GeV if $\sin^2 2 \theta_{13} \sim 0.04$
   ($\theta_{13} \sim 6^0$). The foreseen increase of proton beam intensity is mandatory.

\end{document}